\documentclass[runningheads]{llncs}
\usepackage[T1]{fontenc}
\usepackage{graphicx}
\usepackage{float}  % Add to preamble
\usepackage{wrapfig}  % Add to preamble
\usepackage{placeins}
\usepackage{csquotes}

\begin{document}
\title{Positioning Generative Artificial Intelligence in STEM Assessment: When to Require, Scaffold, or Restrict Its Use}
\titlerunning{GenAI Governance in STEM Assessment}
\author{
Yizhu Gao\inst{1}\thanks{Corresponding author: yizhu.gao@uga.edu} \and
Zhongzhou Chen\inst{2} \and
Min Li\inst{3} \and
Xiaoming Zhai\inst{1}
}

\authorrunning{Y. Gao et al.}

\institute{
University of Georgia, USA \and
University of Central Florida, USA \and
University of Washington, USA\\
\email{yizhu.gao@uga.edu}
}

%\institute{ \and
%Springer Heidelberg, Tiergartenstr. 17, 69121 Heidelberg, Germany
\maketitle              % typeset the header of the contribution
\begin{abstract}

Generative Artificial Intelligence (GenAI) presents a governance challenge for STEM assessment. Unrestricted GenAI access can enable task outsourcing that undermines the validity of traditional assessments; blanket prohibitions are difficult to enforce, may push use underground, and do little to prepare students for workplaces where GenAI-supported workflows are increasingly common. This paper addresses this dilemma by proposing a student-focused framework grounded in Evidence-Centered Design (ECD) that specifies when to \textit{restrict}, \textit{scaffold}, or \textit{require} GenAI use in STEM assessment. The framework extends existing AI-use taxonomies by providing systematic decision rules that link target constructs, evidence requirements, and task characteristics to appropriate governance regimes. We argue that \textit{restrict} is warranted when GenAI threatens construct-relevant evidence for unaided proficiency, particularly for foundational knowledge, canonical representations, and routine procedural skills. \textit{Scaffold} is warranted when bounded GenAI support can reduce peripheral demands (e.g., routine computation or data handling) or when evidence rules for open GenAI collaboration are not yet defensible, thereby preserving interpretability while maintaining task feasibility. \textit{Require} is warranted when the target construct is inherently AI-mediated---such as disciplined human--AI collaboration and discipline--AI literacy---and when tasks are designed to elicit interpretable interaction traces (e.g., prompting, critique, and revision). Using examples from introductory physics, we show how task are under different GenAI use policies. By situating GenAI governance within validity arguments, the framework offers actionable guidance for preserving learning integrity while supporting authentic preparation for AI-enabled professional environments.

\end{abstract}
\section{Introduction}
Generative Artificial Intelligence (GenAI) is reshaping STEM assessment and has prompted debate on how to govern its use within evaluation practices. On one hand, GenAI accelarates answer outsourcing by producing solutions that students may copy and submit, diminishing productive struggle and the learning that comes from engaging with assessment tasks \cite{gao2025multimodal}. On the other hand, GenAI can support tool-mediated practices, such as iteratively generating, critiquing, and revising solutions, that increasingly characterize contemporary STEM workplaces \cite{williams2025integrating}. Preserving both learning from assessment and authentic participation in these emerging practices requires clearer GenAI governance that specifies when its use should be restricted, scaffolded, or required.

Recent work has begun to articulate frameworks for GenAI integration in educational assessment. For example, Furze et al.~\cite{furze2024ai} proposed the AI Assessment Scale, a five-level scheme that specifies permissible uses of GenAI tools in assessment tasks, ranging from \enquote{no AI} to \enquote{full AI}. The scale functions as a communication tool that helps educators define and convey AI-use rules (e.g., ``You must not use AI at any point during the assessment.''). Building on this work, Perkins et al.~\cite{perkins2024ai} refined these levels to \textit{No AI, AI planning, AI collaboration, Full AI,} and \textit{AI exploration}, further illustrating the range of roles AI can play in assessment. However, these taxonomies primarily classify what GenAI may be allowed to do and offer limited guidance on how educators can analyze tasks, select an appropriate level, and adapt or redesign tasks when needed. For example, ``AI collaboration'' implies that students should critically evaluate and revise AI-generated outputs, yet existing frameworks provide little guidance on what tasks can elicit such collaboration or how this depends on the mode of GenAI access (e.g., full-functionality vs.~restricted/guardrailed).

To address these gaps, we propose a student-centered framework for governing GenAI use in STEM assessment, grounded in Evidence-Centered Design (ECD). Rather than treating GenAI governance as a post hoc policy choice, the framework positions GenAI as an explicit design variable within the assessment argument and provides a principled procedure for mapping tasks to governance decisions. Specifically, we show how GenAI access reshapes (a) the \textit{student model}, (b) the \textit{evidence model}, and (c) the \textit{task model}. Using introductory physics tasks as illustrative cases, we demonstrate how this construct-, evidence-, and task-driven analysis supports transparent decisions about when GenAI should be \textit{required}, \textit{scaffolded}, or \textit{restricted}, and how tasks can be redesigned to maintain interpretability when AI use is unavoidable. We conclude by outlining the framework's contributions for assessment design and validity in the era of GenAI, including its implications for defining and measuring emerging human--AI collaborative competencies.

\section{An ECD Lens on GenAI Governance in STEM Assessment}
\begin{figure}
    \centering
    \includegraphics[width=1\linewidth]{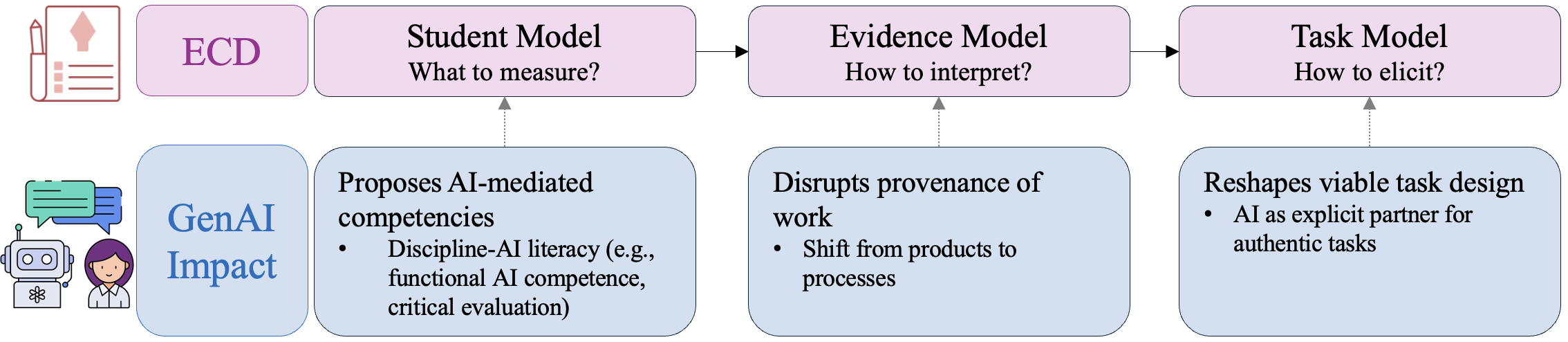}
    \caption{The Three ECD Models and GenAI Impacts}
    \label{fig:ecd-genai}
\end{figure}

ECD is an approach to constructing educational assessments in terms of evidentiary arguments \cite{mislevy2003brief}. It begins with explicit claims about what learners should know and do (\textit{student model}), specifies the evidence needed to support those claims (\textit{evidence model}), and then designs tasks and scoring rubrics to elicit and evaluate that evidence under interpretable conditions (\textit{task model}) (see Figure \ref{fig:ecd-genai}). ECD is especially well suited for analyzing GenAI in assessment because it makes clear where validity can be strengthened or threatened when GenAI is part of the assessment context. Treating GenAI as a design variable prompts systematic analysis and supports decisions to govern its use. We summarize the three core ECD models and analyze how GenAI access affects each one below.

\subsection{Student, Evidence, and Task Models}

\subsubsection{Student Model} 
Student model defines the latent proficiency space targeted by an assessment (e.g. knowledge, skills, strategies) and specify how these proficiencies are structured and related (e.g., skill dependencies, multidimensionality) \cite{mislevy2004case}. In STEM settings, student models are typically anchored in established cognitive and performance frameworks that specify the competence to be elicited and interpreted. Common anchors include Bloom’s taxonomy for characterizing cognitive demand \cite{bloom1956handbook}, national framework documents and standards that define disciplinary learning goals \cite{national2013next}, and career-readiness or 21st-century skills frameworks that emphasize transferrable practices \cite{hilton2012education}. 

\subsubsection{Evidence Model} 
Evidence model specifies how to update beliefs about student-model variables using observable information extracted from learners' work products elicited by assessment tasks \cite{mislevy2003brief}. They comprise (a) \textit{evidence rules} that map work products to observable indicators and (b) a \textit{measurement model} that links those indicators to the targeted student-model variables, updating the inferred likelihood that a learner possesses particular knowledge, skills, or misconceptions. In STEM contexts, work products commonly arise from tasks such as explanation or modeling activities, and laboratory investigations, which elicit students' solutions, representations, and written or oral reasoning.

\subsubsection{Task Model} 
Task model describes how to design assessment situations that will elicit the evidence needed for the evidence model \cite{mislevy2003brief}. It defines the key features of a task—such as the materials presented to the learners and the types of work products the learners generate in response. Importantly, a task model is not a single task; it is a template for a family of tasks that share the same evidentiary purpose. Specific tasks are created by instantiating the model: selecting or authoring the presentation materials and assigning values to the task-model variables that control task features and difficulty. In STEM assessment contexts, task models often instantiate multiple-choice or multiple-select items that vary the stem, representations (e.g., graphs, diagrams, tables), and distractor structure to elicit specific misconceptions and reasoning patterns.

\subsection{How GenAI Reshapes Student, Evidence, and Task Models}

\subsubsection{Impacts of GenAI on Student Model} 
GenAI is reshaping how learners engage in disciplinary work \cite{zhai2025dail}. In higher education, many students report using GenAI to complete or support course-related tasks \cite{baek2024chatgpt}. Accordingly, researchers argue that student model should account for AI-mediated competencies alongside domain knowledge to characterize learners' capacity for strategic GenAI use \cite{zhai2025dail,dera2025developing}. One emerging construct that captures these competencies is discipline-AI literacy, defined as ``the integrated capacity to understand, apply, and critically reflect on AI in authentic disciplinary practices'' \cite{zhai2025dail}. To operationalize discipline-AI literacy, Zhai \cite{zhai2025dail} proposed a six-component framework, including conceptual knowledge of AI in the discipline, practical tool use, data and computational literacy, critical evaluation, ethical and societal understanding, and collaborative and reflective disposition. Complementing this construct-level perspective, Dera \cite{dera2025developing} synthesized case studies on AI literacy in science and engineering education, noting recurring emphases on validating GenAI responses, prompt-oriented instruction, foundations of GenAI, and evaluation of AI-generated content. Taken together, these studies motivate a focused set of competencies that are particularly relevant to STEM learners' interactions with GenAI: \textit{critical evaluation and reasoning}, \textit{functional tool-use competence}, and \textit{collaborative and reflective disposition}. These competencies extend beyond general AI literacy to capture how learners mobilize domain knowledge to interrogate GenAI outputs, make principled judgments about their quality, and productively integrate GenAI into authentic disciplinary work through iterative collaboration.

\subsubsection{Impacts of GenAI on Evidence Model} 
GenAI complicates the evidence model by undermining evidence rules -- the procedures that map student work products to observable variables -- because GenAI can obscure the provenance of final products. For example, when an evidence rule codes the correctness of a physics response as a binary score, that score is no longer interpretable as a direct indicator of the targeted construct if the final solution was generated or substantially shaped by GenAI. The work product remains observable, but the inferential link from product features (e.g., correctness, explanation quality) to student proficiency becomes ambiguous, particularly when high-quality outputs are easiest to outsource. This ambiguity is not entirely new---outsourcing of STEM problem solving has long challenged assessment validity (e.g., via homework-help platforms)---but GenAI markedly intensifies the problem because contemporary models can solve many conventional textbook-style items with passing or better performance. For example, Kortemeyer \cite{kortemeyer2023could} found that ChatGPT could narrowly pass a calculus-based introductory physics course when graded on representative course assessments. Recent systems have also demonstrated near–gold-medal performance on Olympiad-level mathematics under controlled evaluation settings \cite{jian2025loca}, and the latest models such as GPT-5 are reported to achieve (over 80\%) accuracy on standard multimodal assessment items. In such contexts, a correct numerical answer or textual explanation is no longer sufficient evidence for inferring students’ reasoning. This shifts the focal observation from end products that GenAI can readily generate toward products and behaviors that are more difficult to outsource, and from what answer was produced to how it was produced within the assessment context. Accordingly, evidence models should incorporate process-based indicators---such as revision trajectories, intermediate representations, and the timing and sequencing of steps---that preserve a tighter link between observed performance and student proficiency.

\subsubsection{Impacts of GenAI on Task Model} 
GenAI reshapes what assessment tasks are viable and how they should be designed. By enabling generic systems to solve many traditional text-based, static problems, it increases the risk of outsourcing and weakens these items’ validity as measures of individual student thinking. This shift motivates task designs that are richer and more contextualized, and that treat GenAI as an explicit element of the prompt, interaction, or toolset. In addition, higher-order constructs are most appropriately assessed through open-ended, authentic performances \cite{wiggins1991teaching}, yet assessment systems often rely on closed-ended proxies because complex performances can introduce construct-irrelevant variance and reduce interpretability \cite{messick2013alternative}. GenAI can help relax this constraint by scaffolding multi-step work while systematically eliciting and logging intermediate products that provide targeted evidence aligned with intended claims. This expanded design space supports tasks that were previously too time-consuming or demanding—particularly in STEM, where students can engage with real-world datasets and messy phenomena rather than over-simplified synthetic versions. Related “guardrailed” approaches similarly emphasize constrained AI support (e.g., bounded hints or feedback) that aids peripheral skills without revealing target solutions \cite{kapoor2025exploring,liffiton2023codehelp}; in this regime, GenAI functions like a calculator in proof or a lab manual in experimentation, supporting the work while students still demonstrate the core reasoning.

\section{Positioning Generative AI in STEM Assessment}

\begin{figure}
    \centering
    \includegraphics[width=1\linewidth]{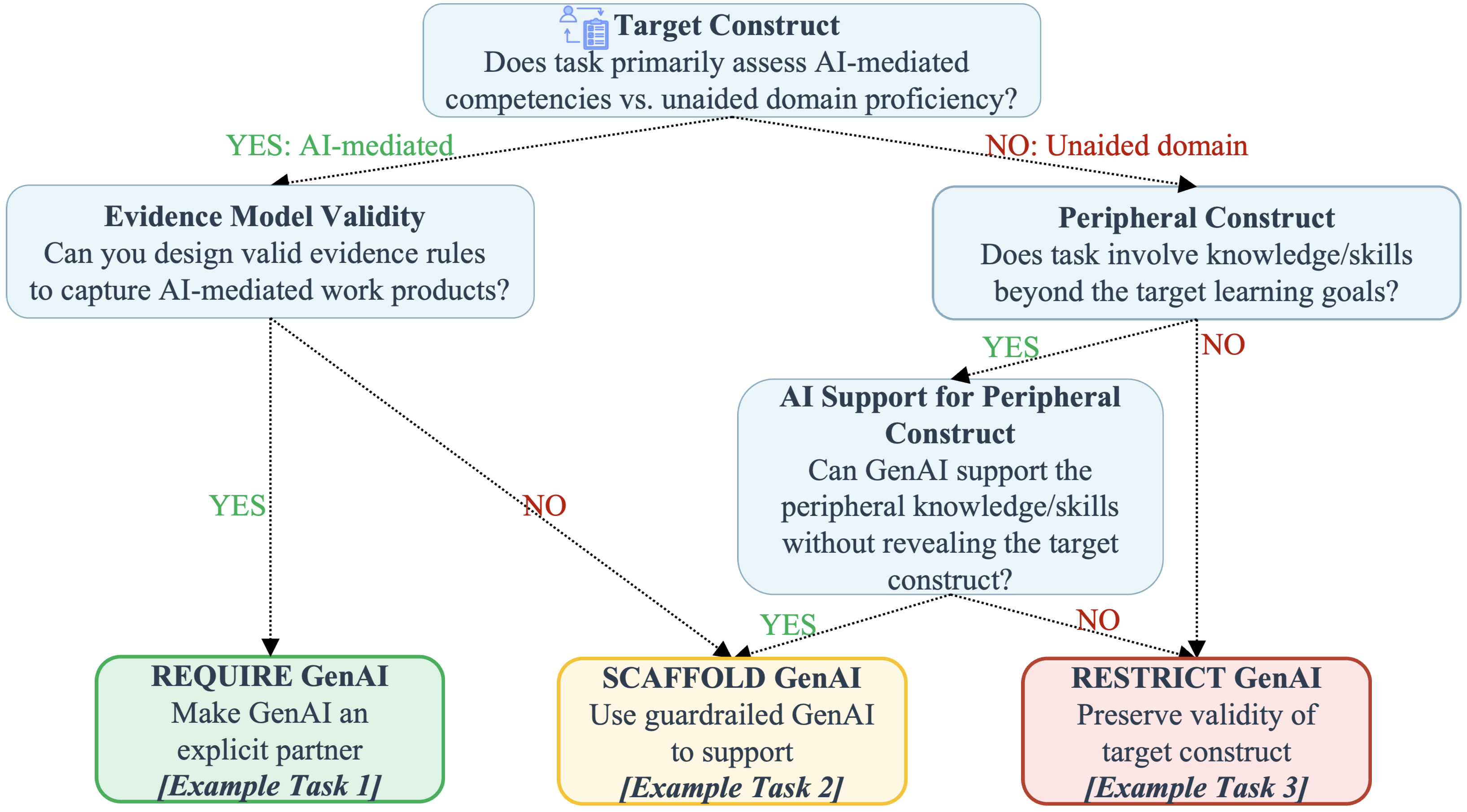}
    \caption{Decision Framework for GenAI Governance in STEM Assessment}
    \label{fig:framework}
\end{figure}

Building on how GenAI affects student, evidence, and task models, we propose a decision framework for answering “Given a particular assessment task, when should GenAI be required, scaffolded, or restricted?” (see Fig. \ref{fig:framework}). The framework follows a sequential decision process. We begin with the target construct in the student model: Does the task primarily assess AI-mediated competencies or unaided domain proficiency? If the target is AI-mediated competence, we next evaluate evidence-model validity: Can the assessment specify evidence rules that support defensive inferences from AI-involved work products? When such evidence rules are feasible, GenAI should be required, positioning it as an explicit partner that is integral to the construct. When valid attribution is not yet achievable, GenAI use should be scaffolded (i.e., constrained and structured) to preserve interpretability. If the target is unaided domain proficiency, the framework shifts to the task model by asking whether the task includes peripheral constructs beyond the core learning goals (e.g., language production, formatting, or routine computation). When peripheral demands are present, GenAI may be scaffolded if it can support those demands without revealing the target construct. If bounded support is not possible—or if the task does not involve peripheral constructs and GenAI would enable outsourcing of the core reasoning—then GenAI must be restricted to preserve the validity of the intended inference. We illustrate each governance with examples from introductory physics.

\subsection{Require GenAI}

\textit{Require GenAI} denotes an assessment governance regime in which GenAI use is mandatory and treated as an explicit task partner because the target construct is inherently AI-mediated. In this regime, students are evaluated on their capacity to engage in interpretable human--AI collaboration—such as prompting, critiquing, verifying, and revising—rather than on unaided domain performance. In the governance framework (Figure.~\ref{fig:framework}), \textit{Require} is warranted when (a) the construct definition centers on AI-mediated competencies (e.g., discipline--AI literacy) and (b) the assessment can specify evidence rules that support defensible inferences from AI-involved work products. Under these conditions, GenAI should be required with full functionality, and tasks should be designed to elicit multi-step collaborative processes rather than a single final answer.

A core implication is that \textit{Require GenAI} tasks should not be satisfiable through a “single-shot” copy-and-submit workflow. Instead, task design should create principled reasons for students to \emph{monitor}, \emph{interrogate}, and \emph{revise} GenAI outputs across multiple iterations, using disciplinary criteria to decide whether to accept, modify, or override AI suggestions. This rationale aligns with the task-contingent nature of contemporary GenAI performance—the “jagged technological frontier” \cite{dell2023navigating}—in which models can perform impressively on some problems yet fail unpredictably on superficially similar ones. Related work argues that such brittleness motivates human-in-the-loop strategies for managing error and uncertainty in consequential settings \cite{han2025general,ju2025collaborating}. These results suggest that \textit{Require GenAI} tasks should be constructed so that initial AI responses are plausible yet diagnostically imperfect, thereby eliciting students’ evaluative and corrective work as observable evidence of AI-mediated competency.

\begin{wrapfigure}{r}{0.5\textwidth}  % r = right side, 0.5 = 50% width
    \centering
    \vspace{-5pt}
    \includegraphics[width=0.42\textwidth]{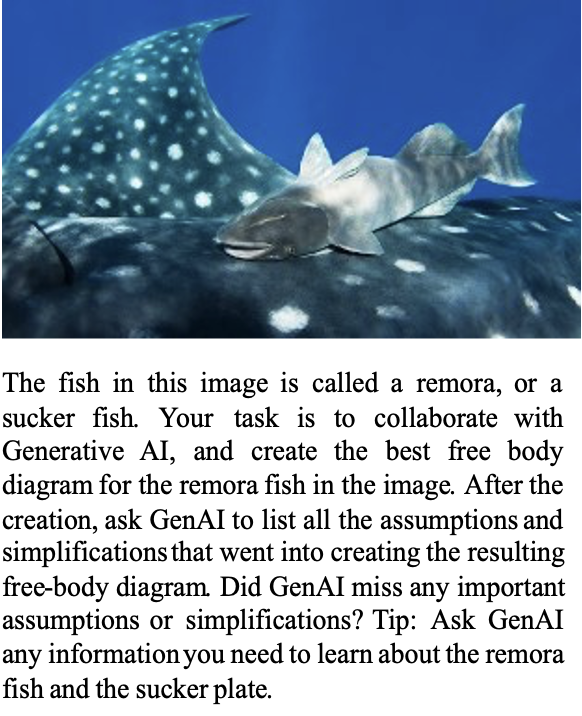}
    \caption{Example Require-GenAI Task}
    \label{fig:remora}
    \vspace{-5pt}
\end{wrapfigure}

STEM domains—particularly physics—provide a strong context for prototyping \textit{Require GenAI} tasks because they offer (i) representational complexity and (ii) principled criteria for verification. Physics problem solving is inherently multimodal, requiring students to coordinate mathematical relationships with diagrams, units, and physical interpretations \cite{hestenes1987toward}. Moreover, physics relies on relatively stable ``world-model'' constraints (e.g., conservation principles, force laws, idealizations) that can be used to validate or falsify candidate solutions. These features are useful because current GenAI systems often struggle with visual/spatial reasoning and with maintaining coherent physical constraints across representations, and they may also produce brittle performance in atypical contexts that deviate from conventional textbook formats \cite{wang2024examining}. Require GenAI tasks can therefore leverage disciplinary constraints to make AI errors identifiable and correctable, foregrounding the student’s role in critique and revision.

Figure~\ref{fig:remora} illustrates a \textit{Require GenAI} task in which students construct a free-body diagram of a remora fish: a fish that attaches itself onto a larger fish via a suction cup on its head. The task is designed to elicit human--AI collaboration through two mechanisms. First, producing a correct free-body diagram requires translating contextual information into a spatially organized representation, creating opportunities for GenAI to generate incomplete or mis-specified forces that students must diagnose. Second, “Remora fish” (or fish of any kind) are seldomly if ever used as problem context in a college introductory level Newtonian Mechanics textbook, making it unlikely for the AI to generate a "standard" solution from its training data. Preliminary probes with state-of-the-art GenAI systems (e.g., GPT-5, Gemini Pro 2.5, Microsoft Copilot, Grok 4.0) resulted in a high probability of producing partially incorrect or poorly justified free body diagrams. The basic concepts (forces, free body diagrams) are grounded in introductory physics context, whereas the task prompts students to acquire and synthesize new information such as the operation of suction cups under water and remora physiology. This makes the task an appropriate probe of human-AI collaboration: students must use disciplinary knowledge to query the AI, combine existing and new knowledge to interrogate and revise its outputs, and construct a coherent, well-supported solution that goes beyond simply using what the AI initially produces. The task requires students to interact with the AI, which enables process-focused evaluation of human-AI collaboration.

\subsection{Scaffold GenAI}

%\textit{Scaffold GenAI} denotes an assessment governance regime in which GenAI use is \emph{permitted but deliberately constrained} so that AI assistance supports task engagement without undermining the interpretability of evidence for the target construct. In contrast to \textit{Require}, which treats GenAI collaboration as the construct itself, \textit{Scaffold} positions GenAI as a \emph{bounded support} whose functionality, timing, and outputs are shaped by explicit evidence rules. As shown in Fig.~\ref{fig:framework}, scaffolding is warranted under three closely related design rationales.

\textit{Scaffold GenAI} denotes an assessment governance regime in which GenAI use is permitted but deliberately constrained to preserve the validity and interpretability of evidence. In the decision framework (Fig.~\ref{fig:framework}), scaffolding is warranted under two principled conditions: (1) when the target construct is AI-mediated but the evidence model is not yet sufficiently developed to support defensible inferences from open-ended GenAI interaction, and (2) when the target construct is unaided domain proficiency but the task involves peripheral knowledge or skills that can be supported by GenAI without revealing the construct of interest. In both cases, scaffolding functions as an evidence-centered control mechanism that specifies \emph{how} GenAI may contribute (e.g., functionality and admissible outputs), rather than simply whether it is allowed.

When the target construct is AI-mediated but the task can be fully outsourced to GenAI, unrestricted access undermines interpretability: students can obtain polished solutions with minimal engagement, yielding interaction traces that reflect tool capability more than the intended human–AI competency. In this setting, restricted GenAI is necessary to preserve evidentiary value by constraining what the system can contribute (e.g., targeted hints, or hypothesis generation without full solutions). These constraints force learners to externalize their own reasoning and make their use of AI diagnostically observable, while still leveraging GenAI as a partner. From an ECD perspective, restricted GenAI functions as an evidentiary safeguard: it aligns task affordances with the current evidence model, reduces construct-irrelevant variance from wholesale outsourcing, and supports more reliable scoring of AI-mediated performance.

When the construct targets unaided domain proficiency, scaffolding is warranted if task completion depends on peripheral demands (e.g., context clarification, representation translation, background information retrieval, or routine technical procedures) that are not central to the intended claims. In this case, GenAI can improve accessibility and feasibility by supporting ancillary processes while being constrained from producing core reasoning steps or final solutions. Such guardrails are essential when unrestricted GenAI access would enable solution substitution and therefore contaminate evidence for domain competence. Accordingly, scaffolds may be implemented through constrained prompts or tool use restricted to non-solution-generating functions.

Across both conditions, scaffolded GenAI preserves construct validity by maintaining explicit control over what counts as admissible evidence. Under scaffolded conditions, the evidentiary focus shifts away from AI-generated text and toward students' evaluative work around it: how they interpret AI feedback, whether they appropriately reject or revise AI suggestions using disciplinary criteria, and how they integrate AI-provided information into coherent reasoning without outsourcing core decisions. In this way, \textit{Scaffold GenAI} supports learning and task feasibility while guarding against construct leakage and premature claims about AI-mediated competency.

\begin{wrapfigure}{r}{0.68\textwidth}  % r = right side, 0.5 = 50% width
    \centering
    \vspace{-5pt}
    \includegraphics[width=0.6\textwidth]{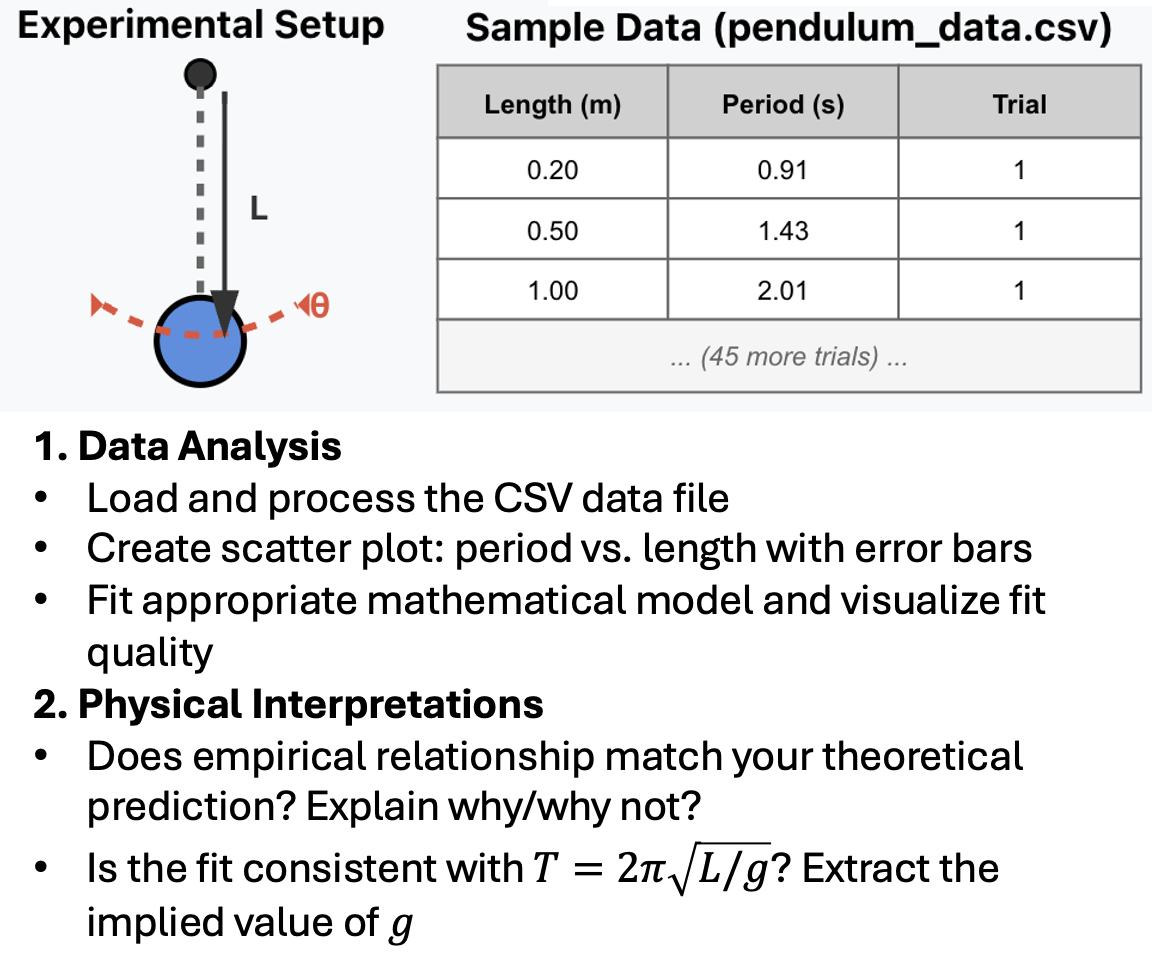}
    \caption{Example Scaffold-GenAI Task}
    \label{fig:pendulum}
    \vspace{-15pt}
\end{wrapfigure}

Figure~\ref{fig:pendulum} illustrates a \textit{Scaffold GenAI} task in which students analyze experimental data on pendulum motion to model the relationship between string length and oscillation period. The task is designed to elicit human--AI collaboration through two mechanisms. First, students are required to make and justify a physics-based prediction of the functional relationship between period \(T\) and length \(L\) using principles of simple harmonic motion, establishing an evidentiary anchor that cannot be outsourced to GenAI. Second, GenAI use is intentionally restricted to technical support functions---such as processing the CSV file, generating plots with error bars, and fitting a specified mathematical model---while all physical interpretation and evaluation must be completed by the student. 

\subsection{Restrict GenAI}
\textit{Restrict GenAI} denotes an assessment governance regime in which GenAI use is prohibited or technically blocked because the target construct is \emph{unaided domain proficiency} and AI assistance would compromise the validity of inferences about that construct. In the decision framework (Fig.~\ref{fig:framework}), \textit{Restrict} is warranted when (a) the focal claims concern students’ independent disciplinary reasoning and (b) GenAI support cannot be cleanly confined to peripheral demands without revealing, substituting for, or strongly cueing the core competencies being assessed. Under these conditions, restricting GenAI preserves the attribution of observed performance to the student and maintains an interpretable mapping from task performance to claims in the student model.

A core implication is that \textit{Restrict GenAI} tasks are those for which GenAI can readily provide solution pathways, canonical representations, or final explanations that are functionally indistinguishable from competent student work. In such cases, even limited, delayed, or ``hint-style'' access may introduce construct leakage: students may infer governing principles, key intermediate steps, or representational structure from AI outputs in ways that are difficult to detect and therefore difficult to incorporate into defensible evidence rules. From an evidence-centered design perspective, restriction is not a normative stance against GenAI; rather, it is a validity-driven governance decision used when AI assistance would introduce construct-irrelevant variance and undermine interpretability of the evidence model.

\textit{Restrict GenAI} most commonly applies to assessments designed to establish a baseline of foundational proficiency, including conceptual recall, fluency with canonical representations, and routine application of well-defined relationships in standard contexts (e.g., introductory ``gateway'' exams, placement tests, and high-stakes summative quizzes). In these assessments, the evidence model typically relies on correctness, clarity, and completeness of unaided responses (selected-response, short-answer, or tightly constrained constructed responses), and the task model is correspondingly static and structured. Allowing GenAI in these settings would shift performance from ``what the student can do'' to ``how well the student can outsource,'' weakening the intended inference.

\begin{wrapfigure}{r}{0.45\textwidth}  % r = right side, 0.5 = 50% width
    \centering
    %\vspace{-5pt}
    \includegraphics[width=1\linewidth]{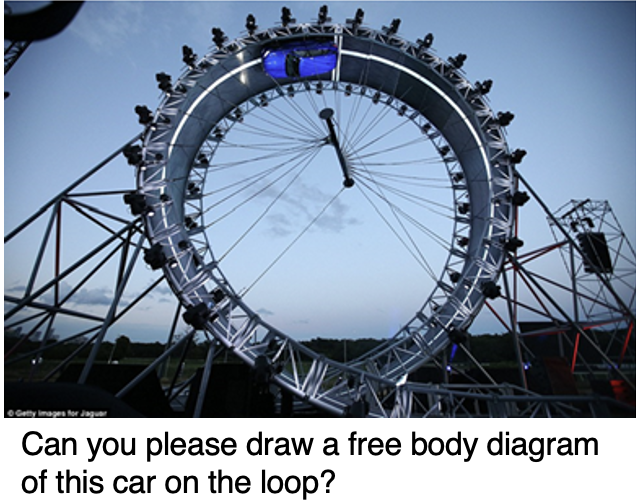}
    \caption{Example Restrict-GenAI Task}
    \label{fig:car}
    \vspace{-15pt}
\end{wrapfigure}

Under restricted conditions, admissible evidence is limited to unaided student work products—such as written explanations, calculations, diagrams, or oral reasoning—collected under conditions that minimize tool-mediated assistance. While this regime does not measure AI-mediated competencies, it provides an essential reference point for interpreting performance under \textit{Scaffold} or \textit{Require} regimes within a broader assessment system.

Figure~\ref{fig:car} illustrates a \textit{Restrict GenAI} task. In the roller-coaster loop problem, students view an image of a car at the top of a vertical loop and are asked to draw a free-body diagram. The student model targets conceptual understanding of forces in circular motion (e.g., weight and normal force) and representational skill in constructing canonical force diagrams. The task model specifies a static multimodal prompt (image plus text) and a constrained diagrammatic response. The evidence model focuses on diagram features such as the inclusion and labeling of relevant forces, their directions, and the coherence of the representation. Because GenAI can often generate a canonical free-body diagram (or force list) that closely matches proficient student work, access to GenAI would risk construct leakage by cueing the representational structure and key components. For high-stakes summative use, we therefore recommend restricting GenAI. For formative use, however, carefully constrained scaffolds (e.g., prompting students to justify each force they add, or providing feedback on common omissions without supplying the full diagram) can support learning while preserving interpretable evidence, especially when revision traces are captured as process data.

\section{Discussion}
\subsection{Positioning the Framework Relative to Prior Work}

Recent frameworks for GenAI in education and assessment have largely emphasized policy guidance, ethical principles, and task-level rules for acceptable use, often framing governance in terms of what should be permitted or prohibited (e.g., allowed vs.\ disallowed tools; acceptable vs.\ unacceptable assistance; \cite{holmes2023guidance}). These contributions have been essential for establishing early norms and guardrails, but they often treat GenAI governance as a constraint applied \emph{after} assessment design, rather than as a decision that follows from what an assessment claims to measure and what evidence is required to support those claims.

The framework proposed here instead treats GenAI governance as a \emph{design variable} embedded within the logic of evidence-centered design (ECD; \cite{mislevy2003brief}). Governance decisions are derived from the student model, evidence model, and task model that together define the intended evidentiary argument. Specifically, governance follows from whether GenAI use is central to the target competency, peripheral to it, or detrimental to the interpretability of evidence. This construct-first orientation provides a principled account of why the same GenAI tool may be restricted in one task, scaffolded in another, and required in a third, depending on how GenAI availability reshapes task demands and the meaning of observed performance \cite{hsiao2023developing,cheng2024evidence}.

Beyond this shift in framing, the framework contributes two clarifications that are often conflated in existing discussions. First, it distinguishes \emph{AI-mediated competencies} from \emph{unaided domain proficiency}, enabling assessment designers to specify when GenAI use is part of the construct rather than a confound to be controlled. Second, it elevates \emph{scaffolding} as an intentional governance strategy with distinct validity-based rationales (e.g., supporting peripheral processes or compensating for underspecified evidence rules), rather than treating it as a compromise between allowance and restriction. Collectively, these contributions position the framework as a validity-oriented decision aid that complements policy and ethics guidance by translating high-level principles into defensible task-level governance choices.

\subsection{GenAI Plays Roles in Human-AI Collaboration Tasks}

Tasks that \textit{require} GenAI constitute a distinct assessment class in which human--AI collaboration is itself the target of measurement. In these tasks, the aim is not to evaluate whether students can produce correct solutions independently, nor merely whether they can use AI efficiently, but whether they can collaborate with GenAI in ways that are disciplined, critical, and reflective to solve domain-relevant problems. Accordingly, Require-GenAI tasks operationalize AI-mediated competency as an assessable construct and treat GenAI interaction as admissible evidence rather than as contamination within the evidentiary argument \cite{mislevy2003brief,cheng2024evidence,zhai2025dail}.

A defining design feature is that GenAI contributions are \emph{treated as uncertain}---often incomplete, under-justified, or plausibly flawed---thereby creating principled opportunities for learners to exercise judgment. GenAI functions as a fallible collaborator whose outputs must be interrogated, revised, triangulated, or rejected using disciplinary reasoning. In this context, the correctness of the final product is often secondary to the quality of the interaction process, including how learners formulate prompts, specify constraints, request evidence, identify gaps or errors, integrate external information, and justify decisions about when to rely on or override AI suggestions \cite{veldhuis2025critical,li2025assessment}.

Because the construct centers on collaboration, Require-GenAI tasks also shift what counts as evidence. Valid evidence extends beyond final artifacts to include interaction traces such as prompt iterations, critique statements, counterexamples, alternative representations, verification moves, and reflective explanations of collaborative choices. Such traces can provide diagnostic insight into learners' ability to manage AI uncertainty, maintain disciplinary grounding in the presence of plausible but flawed outputs, and coordinate human reasoning with AI contributions over multiple turns \cite{cheng2024evidence,li2025assessment}.

Require-GenAI tasks also clarify the relationship between discipline--AI literacy and domain expertise. Effective collaboration presupposes a threshold level of disciplinary knowledge; without it, students cannot evaluate GenAI outputs meaningfully or apply disciplinary criteria to regulate tool use. Thus, Require-GenAI tasks are best interpreted alongside restricted and scaffolded tasks that establish baselines for unaided proficiency and controlled AI use \cite{zhai2025dail,mislevy2003brief}. Taken together, Require-GenAI tasks treat human-AI collaboration as a first-class assessment target and make explicit the forms of process evidence needed to support defensible claims about AI-mediated disciplinary work.

\begin{credits}

\subsubsection{\discintname}
The authors have no competing interests to declare that are
relevant to the content of this article. 
\end{credits}
\bibliographystyle{unsrt}
\bibliography{references}

@article{mislevy2003brief,
  title={A brief introduction to evidence-centered design},
  author={Mislevy, Robert J and Almond, Russell G and Lukas, Janice F},
  journal={ETS Research Report Series},
  volume={2003},
  number={1},
  pages={i--29},
  year={2003},
  publisher={Wiley Online Library}
}

@article{mislevy2004case,
  title={The Case for an Integrated Design Framework for Assessing Science Inquiry. CSE Report 638.},
  author={Mislevy, Robert J},
  journal={Center for Research on Evaluation Standards and Student Testing CRESST},
  year={2004},
  publisher={ERIC}
}

@book{hilton2012education,
  title={Education for life and work: Developing transferable knowledge and skills in the 21st century},
  author={Hilton, Margaret L and Pellegrino, James W},
  year={2012},
  publisher={National Academies Press}
}

@article{zhai2025dail,
  title={DAIL: Discipline-Based Artificial Intelligence Literacy},
  author={Zhai, Xiaoming},
  journal={Available at SSRN 5745703},
  year={2025}
}

@article{gao2025multimodal,
  title={A multimodal interactive framework for science assessment in the era of generative artificial intelligence},
  author={Gao, Yizhu and Zhai, Xiaoming and Li, Min and Lee, Gyeonggeon and Liu, Xiaoxiao},
  journal={Journal of Research in Science Teaching},
  year={2025},
  publisher={Wiley Online Library}
}

@article{williams2025integrating,
  title={Integrating artificial intelligence into higher education assessment},
  author={Williams, Andrew},
  journal={Intersection: A Journal at the Intersection of Assessment and Learning},
  volume={6},
  number={1},
  year={2025},
  publisher={Association for the Assessment of Learning in Higher Education}
}

@article{furze2024ai,
  title={The AI Assessment Scale (AIAS) in Australian K--12 Education},
  author={Furze, Leon and Roe, Jasper},
  journal={Current office holders of the Teachers’ Guild},
  pages={17},
  year={2024}
}

@article{perkins2024ai,
  title={The AI Assessment Scale Revisited: A framework for educational assessment},
  author={Perkins, Mike and Roe, Jasper and Furze, Leon},
  journal={arXiv preprint arXiv:2412.09029},
  year={2024}
}

@article{wiggins1991teaching,
  title={Teaching to the (authentic) test},
  author={Wiggins, Grant},
  journal={Developing minds, a resource book for teaching thinking},
  pages={344--350},
  year={1991},
  publisher={ERIC}
}

@incollection{messick2013alternative,
  title={Alternative modes of assessment, uniform standards of validity},
  author={Messick, Samuel J},
  booktitle={Beyond multiple choice},
  pages={59--74},
  year={2013},
  publisher={Psychology Press}
}

@article{kortemeyer2023could,
  title={Could an artificial-intelligence agent pass an introductory physics course?},
  author={Kortemeyer, Gerd},
  journal={Physical Review Physics Education Research},
  volume={19},
  number={1},
  pages={010132},
  year={2023},
  publisher={APS}
}

@article{jian2025loca,
  title={LOCA-R: Near-Perfect Performance on the Chinese Physics Olympiad 2025},
  author={Jian, Dong-Shan and Li, Xiang and Yan, Chen-Xu and Zheng, Hui-Wen and Bian, Zhi-Zhang and Fang, You-Le and Zhang, Sheng-Qi and Gong, Bing-Rui and He, Ren-Xi and Zhang, Jing-Tian and others},
  journal={arXiv preprint arXiv:2511.10515},
  year={2025}
}

@article{bloom1956handbook,
  title={Handbook I: cognitive domain},
  author={Bloom, Benjamin S and Engelhart, Max D and Furst, Edward J and Hill, Walker H and Krathwohl, David R},
  journal={New York: David McKay},
  pages={483--498},
  year={1956}
}

@article{national2013next,
  title={Next generation science standards: For states, by states},
  author={National Research Council and others},
  year={2013}
}

@article{dera2025developing,
  title={Developing Discipline-Specific AI Ethics Literacy in Science and Engineering: A Call for Faculty and Academic Librarian Collaboration},
  author={Dera, Joanne},
  journal={Science \& Technology Libraries},
  pages={1--10},
  year={2025},
  publisher={Taylor \& Francis}
}

@inproceedings{liffiton2023codehelp,
  title={Codehelp: Using large language models with guardrails for scalable support in programming classes},
  author={Liffiton, Mark and Sheese, Brad E and Savelka, Jaromir and Denny, Paul},
  booktitle={Proceedings of the 23rd Koli Calling International Conference on Computing Education Research},
  pages={1--11},
  year={2023}
}

@article{kapoor2025exploring,
  title={Exploring Student Behaviors and Motivations using AI TAs with Optional Guardrails},
  author={Kapoor, Amanpreet and Diaz, Marc and MacNeil, Stephen and Porter, Leo and Denny, Paul},
  journal={arXiv preprint arXiv:2504.11146},
  year={2025}
}

@article{dell2023navigating,
  title={Navigating the jagged technological frontier: Field experimental evidence of the effects of AI on knowledge worker productivity and quality},
  author={Dell'Acqua, Fabrizio and McFowland III, Edward and Mollick, Ethan R and Lifshitz-Assaf, Hila and Kellogg, Katherine and Rajendran, Saran and Krayer, Lisa and Candelon, Fran{\c{c}}ois and Lakhani, Karim R},
  journal={Harvard Business School Technology \& Operations Mgt. Unit Working Paper},
  number={24-013},
  year={2023}
}

@article{hestenes1987toward,
  title={Toward a modeling theory of physics instruction},
  author={Hestenes, David},
  journal={American journal of physics},
  volume={55},
  number={5},
  pages={440--454},
  year={1987},
  publisher={American Association of Physics Teachers}
}

@article{han2025general,
  title={General reasoning requires learning to reason from the get-go},
  author={Han, Seungwook and Pari, Jyothish and Gershman, Samuel J and Agrawal, Pulkit},
  journal={arXiv preprint arXiv:2502.19402},
  year={2025}
}

@article{ju2025collaborating,
  title={Collaborating with ai agents: Field experiments on teamwork, productivity, and performance},
  author={Ju, Harang and Aral, Sinan},
  journal={arXiv preprint arXiv:2503.18238},
  year={2025}
}

@inproceedings{wang2024examining,
  title={Examining the potential and pitfalls of ChatGPT in science and engineering problem-solving},
  author={Wang, Karen D and Burkholder, Eric and Wieman, Carl and Salehi, Shima and Haber, Nick},
  booktitle={Frontiers in Education},
  volume={8},
  pages={1330486},
  year={2024},
  organization={Frontiers Media SA}
}

@article{baek2024chatgpt,
  title={“ChatGPT seems too good to be true”: College students’ use and perceptions of generative AI},
  author={Baek, Clare and Tate, Tamara and Warschauer, Mark},
  journal={Computers and Education: Artificial Intelligence},
  volume={7},
  pages={100294},
  year={2024},
  publisher={Elsevier}
}

@book{holmes2023guidance,
  title={Guidance for generative AI in education and research},
  author={Holmes, Wayne and Miao, Fengchun and others},
  year={2023},
  publisher={Unesco Publishing}
}

@article{hsiao2023developing,
  title={Developing a framework to re-design writing assignment assessment for the era of Large Language Models},
  author={Hsiao, Ya-Ping and Klijn, Nadia and Chiu, Mei-Shiu},
  journal={Learning: Research and Practice},
  volume={9},
  number={2},
  pages={148--158},
  year={2023},
  publisher={Taylor \& Francis}
}

@inproceedings{cheng2024evidence,
  title={Evidence-centered assessment for writing with generative AI},
  author={Cheng, Yixin and Lyons, Kayley and Chen, Guanliang and Ga{\v{s}}evi{\'c}, Dragan and Swiecki, Zachari},
  booktitle={Proceedings of the 14th learning analytics and knowledge conference},
  pages={178--188},
  year={2024}
}

@article{veldhuis2025critical,
  title={Critical Artificial Intelligence literacy: A scoping review and framework synthesis},
  author={Veldhuis, Annemiek and Lo, Priscilla Y and Kenny, Sadhbh and Antle, Alissa N},
  journal={International Journal of Child-Computer Interaction},
  volume={43},
  pages={100708},
  year={2025},
  publisher={Elsevier}
}

@article{li2025assessment,
  title={An Assessment of Human--AI Interaction Capability in the Generative AI Era: The Influence of Critical Thinking},
  author={Li, Feiming and Yan, Xinyu and Su, Hongli and Shen, Rong and Mao, Gang},
  journal={Journal of Intelligence},
  volume={13},
  number={6},
  pages={62},
  year={2025},
  publisher={MDPI}
}

\end{document}